\documentclass{llncs}
\newcommand{\comment}[1]{{}}
\usepackage[table,xcdraw]{xcolor}
\usepackage{multirow}
\usepackage{url}
\usepackage{wrapfig,lipsum,booktabs}
\usepackage{cleveref}
\usepackage{xspace}
\usepackage{graphicx}
\usepackage{rotating}
\newif\ifsubmission
\submissiontrue 

\newif\ifcomments
\commentstrue

\newcommand{\prover}{$\mathcal P$\xspace}
\newcommand{\verifier}{$\mathcal V$\xspace}
\newcommand{\zia}{{\em ZIA}\xspace}
\newcommand{\wuzia}{{\em WUZIA}\xspace}

\begin{document}
\vspace{-6mm}
\title{Walk-Unlock: Zero-Interaction Authentication Protected with Multi-Modal Gait Biometrics}


\author{
Babins Shrestha \and Manar Mohamed \and Nitesh Saxena \\
\textit{\{babins,manar,saxena\}@uab.edu}
}
\institute{
University of Alabama at Birmingham, USA 
}

\maketitle

\begin{abstract}

	\textit{Zero-interaction authentication} (\zia) refers to a form of
	user-transparent login mechanism using which a terminal (e.g., a
	desktop computer) can be unlocked by the mere proximity of an
	authentication token (e.g., a smartphone). Given its appealing usability, \zia
	has already been deployed in many real-world applications. However, \zia
	contains one major security weakness --- unauthorized physical access to the
	token, e.g., during lunch-time or upon theft, allows the attacker to have
	unfettered access to the terminal.

	In this paper, we address this gaping vulnerability with \zia systems
	by (un)locking the authentication token with the user's walking pattern
	as she approaches the terminal to access it. Since a user's walking or
	gait pattern is believed to be unique, only that user (no imposter)
	would be able to unlock the token to gain access to the terminal in a
	\zia session.  While walking-based biometrics schemes have been studied
	in prior literature for other application settings, our main novelty lies in
	the careful use of: (1) \textit{multiple sensors} available on the
	current breed of devices (e.g., accelerometer, gyroscope and
	magnetometer), and (2) \textit{multiple devices} carried by the user
	(in particular, an ``in-pocket'' smartphone and a ``wrist-worn''
	smartwatch), that all capture unique facets of user's walking pattern.  Our
	contributions are three-fold.  \textit{First}, we introduce, design and
	implement \wuzia (``\textit{Walk-Unlock} \zia''), a multi-modal walking
	biometrics approach tailored to enhance the security of \zia systems (\textit{still} with \textit{zero interaction}). \textit{Second}, we
	demonstrate that \wuzia offers a high degree of detection accuracy,
	based on multi-sensor and multi-device \textit{fusion}.
	\textit{Third}, we show that \wuzia can resist active attacks that
	attempt to mimic a user's walking pattern, especially when
	multiple devices are used.

\end{abstract}

\comment{
\category{H.4}{Information Systems Applications}{Miscellaneous}
\category{D.2.8}{Software Engineering}{Metrics}[complexity measures, performance measures]

\terms{Theory}
}


\vspace{-8mm}
\section{Introduction} \label{sec:intro}

\textit{Zero-interaction authentication} (\zia)~\cite{CN02} represents a rapidly emerging
paradigm, in which a verifier device authenticates a prover
device in physical proximity of the verifier while requiring \emph{no
interaction} by the user of the prover device. The user, carrying
the prover, usually just walks towards the verifier and the verifier 
gets unlocked automatically. In this approach, the prover and verifier devices pre-share a
security association, and simply execute a challenge-response based protocol
for the verifier to authenticate the prover.  
   
The zero-interaction requirement is intended to improve the usability of the
authentication process, which may increase the chances of adoption. Indeed,
\zia\ systems are already getting deployed in many real-world application
scenarios.  For example, BlueProximity \cite{blueProximity} 
allows a user to unlock the idle screen lock in her computer merely by
physically approaching the computer while in possession of a mobile phone,
without having to perform any other action, such as typing in a password. Other
\zia systems include: ``Passive keyless entry and start'' systems like
``Keyless-Go''~\cite{keylessGo},
PhoneAuth~\cite{CzeskisCCS12}, and access control systems based on wearable
devices~\cite{iwatch13}.

However, the zero-interactive nature of \zia systems opens up a fundamental
vulnerability --- unauthorized physical access to the prover device, e.g.,
during lunch-time or upon theft, would allow an attacker to have unfettered
access to the verifier device. Since the prover device does not require any
authorization from the user prior to responding to the verifier device in a
\zia authentication session, mere possession of a lost or stolen prover device
is sufficient to gain access to the verifier device.
Since users' personal devices and items (e.g., smartphones or car keys) are
prone to loss or theft, this issue makes the \zia systems inherently weak and
insecure. 
Speaking about statistics, digital trends \cite{digitalTrends} reports
that Americans lost \$30 billion worth of mobile phones in 2011. Moreover, the
trend has been increasing as reported by Lookout \cite{lookoutReport} that 3.1
million Americans consumers were victims of smartphone theft which is
double the number reported in 2012 by Consumer Reports \cite{consumerReports}.

This raises an important research challenge: \textit{how to protect the} \zia
\textit{systems in the face of loss or theft of prover devices, while still keeping the
authentication process transparent and zero-interactive for the user?} In this
paper, we set out to address this challenge by the use of walking or gait
pattern biometrics prior to authorizing a \zia authentication session.  In
other words, the prover device carried by the user will respond to the
authentication session with the verifier device only when it (the prover
device) detects that it is being carried by the legitimate user. As the user
walks towards the verifier device, the prover device first detects the walking
pattern of the user, and only then gets unlocked and responds to the verifier device. Since a
user's walking pattern is believed to be unique, only that user (no imposter)
would be able unlock the prover device to gain access to the verifier device
in a \zia session.  Since the user has to nevertheless walk towards the
verifier device as part of the \zia authentication process, \textit{no additional effort} is
imposed on the user, thereby preserving the zero-interactivity and user-transparency requirement.

While walking-based biometrics schemes have been studied in prior literature
for other application settings (e.g., \cite{gafurov2006biometric,gafurov2009gait,gafurov2007spoof,huang2007gait,johnston2015smartwatch,kumar2015treadmill,mjaaland2010walk,sondrol2005using,stang2007gait}), our main novelty lies in two important aspects: 

\vspace{-2mm}
\begin{enumerate}

\item The use of \textit{multiple sensors} available on the current breed of devices (e.g.,
accelerometer, gyroscope and magnetometer). 

\item The use of \textit{multiple
devices} carried by the user, in particular, an ``in-pocket'' smartphone and a
``wrist-worn'' smartwatch. Each of these devices capture unique physiological and 
behavioral facets of the user's walking
pattern (e.g., phone captures hip movement and watch captures hand movement).

\end{enumerate}  

\smallskip

\noindent \textbf{Our Contributions:}
The primary contributions of this paper are three-fold:

\vspace{-4mm}
\begin{enumerate}

	\item \textit{\underline{Design of a Walking Biometrics Enhanced ZIA System}}: We introduce,
design and implement \wuzia (\textit{``Walk-Unlock} \zia''), a multi-modal walking
biometrics approach tailored to enhance the security of \zia systems against stolen prover devices
(\textit{still} with \textit{zero-interaction}) . Our \wuzia system uses  
an Android smartphone and/or an Android smartwatch to extract walking biometrics
to authorize a \zia authentication session. \wuzia works with a total of 336 features derived
from 8 sensors of each of the 2 devices. 

\item \textit{\underline{Evaluation under Benign Settings and Passive Attacks}}: 
We demonstrate that \wuzia offers a high degree of detection
accuracy, based on multi-sensor and multi-device \textit{fusion}. Our results
show that walking biometrics can be extracted with
a high overall accuracy when using one of the devices (phone or watch), and 
became almost error-free when both devices were used together (i.e., 0.2\% false negatives and 0.3\% false positives on average).
This suggests that \wuzia can be highly accurate in detecting 
a valid user as well as an unauthorized entity who 
(accidentally or deliberately) walks towards the authentication terminal.

\item \textit{\underline{Evaluation under Active Imitation Attacks}}: We show that \wuzia
	can resist active attacks that deliberately attempt to mimic a user's
	walking pattern, including a state-of-the-art \textit{treadmill-based
	attack} \cite{kumar2015treadmill}.  In particular, our results suggest that, especially when using both
	devices (phone and watch), such attacks would become very difficult in practice (4.55\% false positives on average) even when 
	the attacker capabilities are very high.

\end{enumerate}

\vspace{-4mm}


\vspace{-4mm}
\section{Background} \label{sec:background}


In this section, we define a Zero-interaction authentication (\zia) system, 
present the existing threat model for such a system, and then enumerate the design goals of our proposed system. 

\vspace{-3mm}
\subsection{Zero-Interaction Authentication} 
\label{subSec:zia}

A \zia system relies
upon the authentication factor ``something you have''.
A \zia scheme involves a user who carries a prover device (\prover) and needs
to validate her identity to a verifier device (\verifier).  \prover and \verifier 
typically communicate over a short-range wireless communication channel such as Bluetooth. 
\prover and \verifier share a prior security association (shared key $K$) and the messages
between them are encrypted and authenticated. In particular,
a \zia authentication session runs a challenge-response authentication protocol that
authenticates \prover to \verifier. 
That is, \prover sends a random challenge $C$ to \verifier, and \verifier returns back a 
response $R$ which is an authenticated encryption of $C$, in order to prove the 
possession of the shared key $K$.
The user does not need to perform
any explicit action or gestures in the authentication process. Simply walking 
towards \verifier, while carrying \prover, establishes the authentication. 

\vspace{-3mm}
\subsection{Threat Model} \label{subSec:threat}

In \zia threat model, \prover and \verifier are assumed to be honest (i.e.,
uncompromised and non-malicious).  The communication channel between \prover
and \verifier is protected with encryption and authentication tools. 

In a realistic threat model, an attacker should be assumed to be in possession of the \prover
device. The attacker may obtain the \prover device either by stealing it or
via a lunchtime attack \cite{eberz2015preventing}.  In this model, existing
\zia systems are completely broken since the attacker can just access \verifier
by making use of \prover.  

\zia systems are known to be vulnerable to relay attacks.  This is because the
user usually carries \prover and gets verified when she simply comes near to
\verifier over radio-frequency (RF) signals.  A relay attacker's goal is to
relay these RF signals from  \prover to \verifier such that the attacker is
authenticated without possessing \prover.  Security researchers have proposed
various techniques to defend against relay attacks such as using distance time
bounding \cite{BC93,Hancke05,Reid07} or using context information from the
environment \cite{halevi2012secure,TruongPerCom14}.  
As such, the threat model assumes that a relay attack prevention technique has already been deployed.
However, such a technique can not defend against the theft or loss of the
\prover device (this is the vulnerability we aim to address in this paper).  

\vspace{-3mm}
\subsection{Design Goals and Metrics} \label{subSec:goals}

For a \zia scheme that remains secure even under the event of loss or theft of the \prover device, like 
the one proposed in this paper, following design criteria must be considered:

\vspace{-2mm}
\begin{enumerate}
\item \textit{Lightweight}: The scheme should not be memory intensive or
computationally intensive. It should be lightweight in terms of power usage as the battery life is one of the most important factors for user's device usage.
\item \textit{Efficient}: The approach should not incur perceptible delay. Users should
not be required to wait for a long period to get authenticated.
\item \textit{Robust}: The scheme should be robust to errors and attacks. 
The system must authenticate with high probability when an authorized user with 
\prover is authenticating to \verifier while the unauthorized users must be denied 
access to \verifier. It must also be robust towards the active attackers who may intentionally attempt 
to bypass the system (e.g., mimic the user's walking pattern on our proposed scheme).
\item \textit{Transparent \& Zero-Effort:} Since the approach is zero-interactive, the authentication 
should be transparent to the users. The users should not be required to perform additional 
tasks (such as typing passwords/pins) or explicit gestures.
These 
actions may degrade the usability of the system, and reduce chances of adoption.
\end{enumerate}

\vspace{-4mm}
\section{Our Approach: Walk-Unlock ZIA} \label{sec:approach}

To protect the unlocking of \verifier in the face of loss or theft of \prover in a \zia scheme,
we propose to authenticate the user based on a gait-based authentication system.  In
other words, we propose to authenticate the user with her unique walking pattern.
Different categories of sensors are embedded nowadays into smartphones and smartwatches 
such as motion, position and environment sensors.
Android OS, one of the most popular smart device operating systems, provides
APIs to support different categories of these sensors. 
We leverage these sensors, especially motion and
position sensors, to identify that the \prover device is undergoing a
particular activity, in a specific motion and orientation, as if the prover device is
being carried/worn by the legitimate user.  This activity detected by the \prover device
is transparent to the user since it is performed implicitly while the user
walks towards \verifier. 

\begin{figure}[t]
\vspace{-10mm}
\centering
  \includegraphics[width=1\columnwidth]{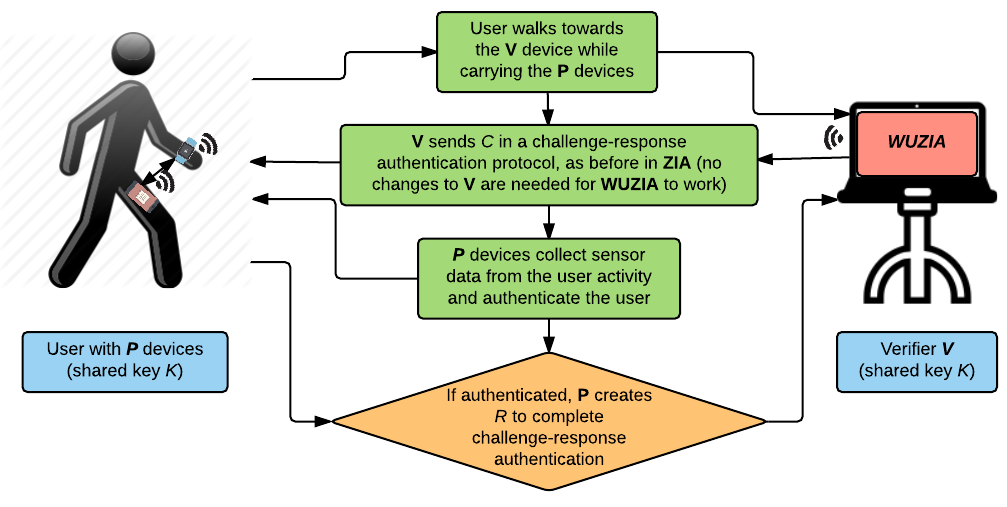}
\vspace{-6mm}
\caption{\wuzia system overview: the \prover devices respond to 
the \verifier device in a challenge-response \wuzia authentication 
only if the \prover devices detect the valid walking pattern of the user. 
In this paper, we consider a smartphone and a smartwatch as the \prover devices. 
}
 \label{fig:processFlowChart}
 \vspace{-4mm}
\end{figure}

While many types of \prover devices may be used to detect the user's walking
activity prior to authorizing a \zia session, in this paper, we capture the walking
biometrics using an ``in-pocket'' device and/or a ``wrist-worn'' device, both
devices having multiple on-board sensors. 
Specifically, in such a walk-unlock \zia (\wuzia) scheme, we aim to authenticate the user in a robust
manner using machine learning classifiers based on data drawn from multiple
sensors from multiple devices such as smartphone (in-pocket) and smartwatch
(wrist-worn).
The \wuzia authentication process has been visualized in Figure
\ref{fig:processFlowChart}.
As shown in Figure \ref{fig:processFlowChart}, \wuzia requires changes only in the \prover devices.
The \verifier device in an existing \zia system is transparent to the authentication process
and requires no modification. Hence, \wuzia can be implemented in traditional \zia
system such as BlueProximity \cite{blueProximity} by just changing the smartphone 
app, without changing the terminal software. 

In our \wuzia system, we use multiple devices, i.e., a smartphone and a smartwatch, to authenticate the
user. However, to analyze the efficiency and robustness of our system systematically, we show:
\vspace{-2mm}
\begin{enumerate}
\item walking pattern extraction using the in-pocket smartphone,
\item walking pattern extraction using the wrist-worn smartwatch, and
\item combination of the above two.
\vspace{-1mm}
\end{enumerate}

The second setting is suitable for situations where the user may leave her phone on the desk 
space or the car dashboard, and will need to be logged in just by using her watch. 
Although currently most of the smartwatches work along with companion devices (smartphones), 
we believe that in the future such devices would be usable as stand-alone devices. 

The threat model of \wuzia is in line with that of \zia (Section \ref{subSec:threat}), except that 
the former aims to be secure even under the adversarial possession of \prover.
Since in the proposed scheme, \prover can be either a smartphone or a smartwatch or both, the attacker may therefore possess only one of the devices or both devices.
After the attacker possesses \prover (one or both devices), it will try to
unlock \verifier. Further, a \wuzia attacker may be active in the sense that it may try
to authenticate itself as the valid user by mimicking the walking pattern of the
user as measured by \prover device(s). We allow such an attacker to observe (and record) the user in an attempt
to imitate the user's walking habits. 

In the \wuzia system, we
assume that a relay attack prevention technique has already been deployed (like in a \zia system).
That is, no relay attacks are possible between \prover and \verifier.  
Similarly,
we assume that no relay attacks are possible between the \prover devices (phone
and watch). Also, we assume that the two \prover devices are securely paired with each other and 
that all communication between them
has been protected with traditional cryptographic mechanisms. 

Given this threat model, in the following sections, we will show that our \wuzia system satisfies all of our
design goals (Section \ref{subSec:goals}), i.e., being lightweight, efficient, robust and transparent.

\vspace{-2mm}
\section{Data Collection: Design and Procedures}

\label{sec:dataCollection}
To develop and evaluate our system for authenticating the users based on their walking pattern, we need to collect the sensors data from the users' smartphones and smartwatches while they are walking. 
We developed a framework that encompasses two Android apps and a web app. 
The web app utilizes Google Cloud Messaging (GCM) to send commands to the smartwatch. One of the Android apps is installed on the smartphone and the other is installed on the smartwatch. 
 


\vspace{-2mm}
\begin{enumerate}
\item \textit{Web App:} We used GCM to send start/stop commands to the smartphone, which upon receiving start/stop recording the sensors data and send start/stop recording trigger to the smartwatch. We created a simple HTML page with a text box to record the user information, a start recording button, and a stop recording button. 
The experimenter first inputs the user information in the text box and hits the start recording 
button when the user starts walking towards \verifier. When the user touches \verifier, the 
experimenter hits the stop recording button.
We used GCM for the purpose of data collection only (in real-life implementation, GCM is not needed).

\item \textit{Smartphone App:} The app on the smartphone waits for the GCM commands. As soon as it receives the GCM start command, it sends a start recording trigger to the smartwatch and starts recording the sensors value. As soon as it receives the GCM stop command, it sends a stop recording trigger to the smartwatch and stops recording the sensors value. 

\item \textit{Smartwatch App:} The app on the smartwatch waits for the smartphone's triggers. Once it receives a start recording trigger, it starts recording the sensors values and keeps on recording until it receives a stop recording trigger. The recorded sensor values by the smartwatch are stored in the smartwatch.

\end{enumerate}

The sensors utilized in our implementation, from both 
the smartphone and the smartwatch, are listed in
Table \ref{table:Device_sensors}.

\begin{table}[]
\centering
\scriptsize
\vspace{-5mm}
\caption{Sensors utilized for walk biometrics.}
\label{table:Device_sensors}
\renewcommand{\arraystretch}{1.2}
\vspace{-2mm}
\begin{tabular}{|l|l|l|}
\hline
\textbf{Sensor Name}     & \textbf{Sensor Type} \hspace{2mm} & \textbf{Description}                     \\ \hline\hline
Accelerometer (A)        & Motion               & The acceleration force including gravity \\ \hline
Gyroscope (Gy)           & Motion               & The rate of rotation                     \\ \hline
Linear Acceleration (LA) & Motion               & The acceleration force excluding gravity \\ \hline
Rotation Vector (R)      & Motion               & The orientation of a device              \\ \hline
Gravity (G)              & Motion               & The gravity force on the device          \\ \hline
Game Rotation (GRV)      & Position             & Uncalibrated rotation vector             \\ \hline
Magnetic Field (M)       & Position             & The ambient magnetic field               \\ \hline
Orientation (O)          & Position             & The device orientation                   \\ \hline
\end{tabular}
\vspace{-5mm}
\end{table}

\comment{
In \zia, the app communicates via RF signal as described in Section \ref{subSec:zia}.
When \prover is near \verifier, \verifier sends a challenge to \prover 
\cite{blueProximity,truong2015using,TruongPerCom14} upon which \prover has 
to reply back with a response.
\wuzia will be triggered when \prover needs to compute this response. 
Once the user
is authenticated, \wuzia allows \prover to relay the message, otherwise the challenge
from \verifier is dropped.
}

For data collection, we recruited 18 students in our 
University through the word of mouth. Among these 
participants, 15 were male while 3 were female. 
To avoid any kind of inconsistency, we used only 
one smartphone (LG Nexus 5 (D820) \cite{nexus5}) and 
one smartwatch (LG G watch R (W110) \cite{LGGWatchR}).
Both devices 
have Android OS version 6.0.1. We conducted the 
experiment following the University's IRB guideline. 
The participants were clearly informed about the 
experiment such as the data being collected, the 
purpose of the experiment, and that they can refuse to 
participate in the middle of the experiment or even 
request to delete their collected data during or after 
the experiment has been conducted. Our University's 
Institutional Review Board approved the project.

After the participants were detailed about the 
experiment, we asked these volunteers to wear the 
smartwatch on their (left/right) hand where they 
normally wear their watch and put the smartphone in 
their (left/right) pocket where they normally put it 
during walking. We asked each volunteer to walk from a 
door to the computer (distance of around 7 meters) as 
if they are trying to log in. The experimenter sent 
the GCM command to the smartphone to start the sensors 
recording when the user started walking. As soon as the 
user touches the keyboard as if the user is trying to log 
into the computer, the experimenter sent another GCM 
command to stop the sensors recording. We noticed that 
some of the users log into the machine standing while 
others sit on a chair before they touch the keyboard. 
One of the participants even placed his phone on the 
desk before he logged into the machine.

We collected the data from these volunteers for a 
period of time ranging from 30 to 60 days based on their 
availability. We asked each user to 
walk for around 7 meters (from the door to the machine) 
for five times each day. We collected the data from 
each user for 10 days resulting in 50 samples of walking 
data from each user.

\vspace{-3mm}
\section{Gait Biometrics Detection: Design and Evaluation} \label{sec:eval}

\vspace{-1mm}
In order to evaluate the performance of the proposed gait biometrics as
an authentication scheme, we utilized the machine learning approach based on
the underlying readings of the motion sensors, and the position sensors from both of the phone and the watch. 

\vspace{-4mm}
\subsection{Preliminaries}
\vspace{-2mm}
\noindent\textbf{Classifier:} In our analysis, we utilized the Random Forest classifier. Random Forest is an ensemble approach based on the generation of many
classification trees, where each tree is constructed using a separate bootstrap
sample of the data. To classify a new input, the new input is run down on
all the trees and the result is determined based on majority voting. Random Forest is efficient, can estimate the
importance of the features, and is robust against noise \cite{maxion2010keystroke}. 
Random Forest outperforms other classifiers including support vector machines which are considered to be the best classifier currently available \cite{caruana2006empirical,liu2013comparison,maxion2010keystroke}.
 
\smallskip
\noindent\textbf{Features:} For each of the used sensor
instances, we calculated the mean, the standard deviation and the range of each of the axis ($X, Y, Z$), the square of each axis ($X^2$, $Y^2$, $Z^2$) and the square root of the sum of squares for that
instance's axes components ($X, Y, Z$) of
all the instances in the sample that corresponds to a single walk instance. Twenty one features are extracted from each of the used sensors, which give us a total of 336 features.

The 336 features or subset of them were used as input to train the classifier to
differentiate a user from other users.  
%
In the classification task, the positive class corresponds to the gait
of the legitimate user and the negative class corresponds to impersonator (other user).
Therefore, true positive (TP) represents the number of times the legitimate
user is granted access, true negative (TN) represents the number of times
the impersonator is rejected, false positive (FP) represents the number of
times the impersonator is granted access and false negative (FN) represents the
number of times the correct user is rejected.

As performance measures for our classifiers, we used false positive, false negative, precision, recall and
F-measure (F1 score), as shown in  
Equation \ref{eqn:precision_recall}.
FP/precision measures the security of the proposed system, i.e., the accuracy of
the system in rejecting impersonators. FN/recall measures the usability of the
proposed system as high FN leads to high rejection rate of the legitimate
users.  F-measure considers both the usability and the security of the system.
To make our system both usable and secure, ideally, we would like to have FP and FN
as close as 0 and recall, precision and F-measure as close as 1. 

\vspace{-1mm}
{\scriptsize{

\begin{equation}
\vspace{-3mm}
\label{eqn:precision_recall}
precision = \frac{TP}{TP+FP}; \hspace{5mm} recall = \frac{TP}{TP+FN}; \hspace{5mm} F\mbox{-}measure = 2* \frac{precision*recall}{precision+recall}\\
\end{equation}
\vspace{-3mm}
}}

%

\begin{table}[t]
\centering
\vspace{-6mm}
\scriptsize
\caption{Performance for the classifier for three different 
categories. The first three rows show the performance of the classifier using all the sensors. The next three rows show the results of using the sensors subset that provides the best average results. The last three rows show the result
of using the best sensors subset for each user.  Highlighted cells emphasize the most interesting results.}
\label{table:ResultsGeneral}
\def\arraystretch{1.7}
\begin{tabular}{ll|c|c||c|c|c|}
\cline{3-7}
                                                           &                     & \textbf{FNR}    & \textbf{FPR}    & \textbf{F-Measure} & \textbf{recall} & \textbf{precision} \\ \cline{3-7} 
                                                           &                     & \multicolumn{5}{c|}{\textbf{Avg (std. dev.)}}                                          \\ \hline \hline
\multicolumn{1}{|l|}{\multirow{3}{*}{\rotatebox[origin=c]{90}{\textbf{Overall}}}} & \textbf{Phone Only}  & 0.058 (0.037)   & 0.068 (0.034)   & 0.937 (0.026)      & 0.942 (0.037)   & 0.934 (0.031)       \\ \cline{2-7} 
\multicolumn{1}{|l|}{}                                     & \textbf{Watch Only} & 0.085 (0.050)   & 0.105 (0.045)   & 0.906 (0.036)      & 0.915 (0.050)   & 0.899 (0.040)  \\ \cline{2-7} 
\multicolumn{1}{|l|}{}                                     & \textbf{Both}       & 0.038 (0.047)   & 0.042 (0.031)   & 0.960 (0.030)      & 0.962 (0.047)   & 0.960 (0.029)      \\ \hline\hline
\multicolumn{1}{|l|}{\multirow{3}{*}{\rotatebox[origin=c]{90}{\textbf{General}}}}    & \textbf{Phone Only} & 0.040 (0.035)   & 0.051 (0.033)   & 0.954 (0.025)      & 0.960 (0.035)   & 0.950 (0.031)      \\ \cline{2-7} 
\multicolumn{1}{|l|}{}                                     & \textbf{Watch Only} & 0.080 (0.049)   & 0.095 (0.043)   & 0.913 (0.030)      & 0.920 (0.049)   & 0.909 (0.038)        \\ \cline{2-7} 
\multicolumn{1}{|l|}{}                                     & \textbf{Both}       & 0.022 (0.027)   & 0.030 (0.027)   & 0.974 (0.021)      & 0.978 (0.027)   & 0.971 (0.025)       \\ \hline\hline
\multicolumn{1}{|l|}{\multirow{3}{*}{\rotatebox[origin=c]{90}{\textbf{\hspace{-3mm} Individual \hspace{-5mm} }}}} & \textbf{Phone Only} 
																				 & 0.018 (0.023)   & 0.036 (0.020)   & 0.973 (0.013)      & 0.982 (0.023)   & 0.965 (0.019)        \\ \cline{2-7} 
\multicolumn{1}{|l|}{}                                     & \textbf{Watch Only} & 0.046 (0.034)   & 0.063 (0.044)   & 0.947 (0.024)      & 0.954 (0.034)   & 0.941 (0.039)      \\ \cline{2-7} 
 
\multicolumn{1}{|l|}{}                                     & \textbf{Both}       & \cellcolor[HTML]{C0C0C0}{0.002 (0.006) }  & \cellcolor[HTML]{C0C0C0} {0.003 (0.008) }  & \cellcolor[HTML]{C0C0C0} 0.997 (0.005)      & \cellcolor[HTML]{C0C0C0} 0.998 (0.006)   & \cellcolor[HTML]{C0C0C0} 0.997 (0.008)       \\ \hline
\end{tabular}
\vspace{-4mm}
\end{table}

\vspace{-3mm}
\subsection{Classification Results}

As mentioned in Section \ref{sec:dataCollection}, we collected data from 18
users. From each user, we collected 50 samples of walking data. We divided the collected data
into 18 sets based on the users' identities (ids).  In order to build a classifier to
authenticate a user based on her gait biometrics, we defined two classes.
The first class contains the walking data from a specific user, and the other class
contains randomly selected walking data from other users.   

The classification results are obtained after running a 10-fold cross validation, and are summarized in Table
\ref{table:ResultsGeneral}. 
The first part of Table \ref{table:ResultsGeneral} shows the results of using all the features 
extracted using sensors from the phone, the watch and both devices. 
We found combining the features from the phone and the watch sensors decreases the false 
negative from 5.8\% in case of only phone, 8.5\% in case of using only watch to 3.8\% and decreases 
the false positive from 6.8\% in case of only phone, 10.5\% in case of using only watch to 4.2\%.

The second part of Table \ref{table:ResultsGeneral} shows the results obtained by finding the 
sensor subset that provides the best overall average. We found that utilizing only  
accelerometer, gyroscope, magnetometer and orientation sensors from phone rather than using 
all phone sensors decreases the false negative and the false positive by around 2\%.
Similarly, using only  accelerometer, gravity, gyroscope, linear acceleration and magnetometer 
sensors from watch instead of using all watch sensors decreases the false positive rate from 
10.5\% to 9.5\% and the false negative rate from 8.5\% to 8.0\%.
Furthermore, we found utilizing only phone accelerometer, phone gyroscope, phone magnetometer, 
phone orientation, watch accelerometer, watch magnetometer, and watch orientation sensors 
improves the classification accuracy (i.e., decrease both the false positive and the false 
negative rate by 1.2\% and 1.6\%, respectively).
These features subset also contained the subset of features which were not correlated to each other. 
We leverage these uncorrelated features to prevent our \wuzia system against a sophisticated 
form of active impersonation attack \cite{kumar2015treadmill}, as we will describe in Section \ref{sec:treadmill}.

Finally, we checked the classification accuracy by selecting for each user the subset of 
sensors that provides the best results. The results of this model are shown in the last three 
rows of Table \ref{table:ResultsGeneral}. 
We found out that the classifier performance improved over 
the previous two models. Moreover, both the average false positive and the average false 
negative rates dropped to around 0\% when we used the best subset from both of the devices.

In summary, the results obtained from the classification models show that the gait biometrics can be detected in a robust manner and thus will serve as an effective method for authenticating the users.
The results show that the fusion of the phone and the watch sensors significantly enhances the performance of detecting the gait biometrics.
This is reflected in very low false positives and false negatives. 
 


\vspace{-2mm}
\section{Resistance to Active Attacks} \label{sec:attack}
\vspace{-2mm}
\subsection{Human Imposter Attack} \label{sec:imposter}
\vspace{-1mm}
In a human-based imposter attack, the adversary tries to manually mimic a victim's walking pattern so that
it can fool the \wuzia system. Our model assumes that 
the attacker already has the physical possession of the \prover devices (phone and/or watch). 
Such kinds of attacks
have been explored in the literature by few researchers \cite{gafurov2007spoof,mjaaland2010walk,stang2007gait}. 
However, most of these works use accelerometer devices (e.g., MR100 wearable sensor) (not a phone or a watch used in our scheme), and these devices are worn on the waist tied to the belt  
\cite{gafurov2007spoof,mjaaland2010walk} or on the limbs near the shoes \cite{stang2007gait}.
Therefore, we analyze how our system will perform when an attacker with similar physical characteristics 
attempts to learn and imitate an individual's walking pattern. 

During the walking biometrics data collection, we recorded videos of eight different users. The attacker (a researcher, serving the role of an expert attacker) 
chose two of the users as victims (we call them $V_1$ and $V_2$) who exhibited the simplest walking pattern or distinctive visible 
characteristics, upon careful visual inspection.  
If the attacker can not succeed in attacking such simplistic walking patterns, then 
it would be harder for the attacker to succeed in attacking more complex walking patterns.  

In our experiment, the attacker watched the video several times so as to learn the feet 
and the hand movement pattern of the user. 
While practicing, the attacker also tried to match the time duration from the start
to the end of the victim's walk, using the video.
After the attacker felt comfortable with the timing and the walking pattern, we
collected the data for the attacker with the \prover devices walking towards the \verifier device. The attacker
was provided the visual feedback while imitating the walk pattern.

\begin{table}[t!]
\centering
\scriptsize
\caption{Performance for the imposter attack on two different victim users for two different types of 
classifier categories. The first three rows show the performance of the imposter attack against the 
classifier trained with the subset that provides best average results, as mentioned in Table \ref{table:ResultsGeneral}. The last three rows show the result of the imposter attack against the classifier
trained with the best subset for the individual victim user. Highlighted cells emphasize the most interesting results.}
\label{tab:ImposterResult}
\def\arraystretch{1.7}
\begin{tabular}{lll|ccclccc|}
\cline{4-6} \cline{8-10}
                                       &                                          & \multicolumn{1}{l|}{} & \multicolumn{2}{c|}{\textbf{Victim $V_1$}}                                         & \multicolumn{1}{c|}{\textbf{Attacker}} & \multicolumn{1}{l|}{} & \multicolumn{2}{c|}{\textbf{Victim $V_2$}}                                         & \multicolumn{1}{c|}{\textbf{Attacker}} \\ \cline{4-6} \cline{8-10} 
                                       &                                          & \multicolumn{1}{l|}{} & \multicolumn{1}{c|}{\textbf{F-measure}} & \multicolumn{1}{c|}{\textbf{FPR}} & \multicolumn{1}{c|}{\textbf{FPR}}      & \multicolumn{1}{l|}{} & \multicolumn{1}{c|}{\textbf{F-Measure}} & \multicolumn{1}{c|}{\textbf{FPR}} & \multicolumn{1}{c|}{\textbf{FPR}}      
\\ \hline \hline 
\multicolumn{1}{|c|}{\multirow{3}{*}{\rotatebox[origin=c]{90}{\bf General}}} & \multicolumn{1}{c|}{\textbf{Phone Only}} & \multicolumn{1}{c|}{}  & \multicolumn{1}{c|}{0.931}   & \multicolumn{1}{c|}{0.100}                    & \multicolumn{1}{c|}{0.000}                 & \multicolumn{1}{c|}{} & \multicolumn{1}{c|}{0.936}             & \multicolumn{1}{c|}{0.100}           & \multicolumn{1}{c|}{0.917}             \\ \cline{2-2} \cline{4-6} \cline{8-10} 
\multicolumn{1}{|c|}{}                  & \multicolumn{1}{c|}{\textbf{Watch Only}}  & \multicolumn{1}{c|}{} & \multicolumn{1}{c|}{0.887} & \multicolumn{1}{c|}{0.100}                        & \multicolumn{1}{c|}{0.909}             & \multicolumn{1}{c|}{} & \multicolumn{1}{c|}{0.935}              & \multicolumn{1}{c|}{0.080}         & \multicolumn{1}{c|}{0.000}                 \\ \cline{2-2} \cline{4-6} \cline{8-10} 
\multicolumn{1}{|c|}{}             & \multicolumn{1}{|c|}{\textbf{Both}}       & \multicolumn{1}{c|}{}  & \multicolumn{1}{c|}{0.980}              & \multicolumn{1}{c|}{0.020}         & \multicolumn{1}{c|}{0.091}             & \multicolumn{1}{c|}{} & \multicolumn{1}{c|}{0.989}             & \multicolumn{1}{c|}{0.000}             & \multicolumn{1}{c|}{0.833}             \\ 
 \hline \hline
\multicolumn{1}{|c|}{\multirow{3}{*}{\rotatebox[origin=c]{90}{\bf Individual}}} & \multicolumn{1}{c|}{\textbf{Phone Only}} & \multicolumn{1}{c|}{} & \multicolumn{1}{c|}{0.970}               & \multicolumn{1}{c|}{0.040}         & \multicolumn{1}{c|}{0.182}             & \multicolumn{1}{c|}{} & \multicolumn{1}{c|}{0.968}              & \multicolumn{1}{c|}{0.060}         & \multicolumn{1}{c|}{0.917}             \\ \cline{2-2} \cline{4-6} \cline{8-10} 
\multicolumn{1}{|c|}{}                  & \multicolumn{1}{c|}{\textbf{Watch Only}}  & \multicolumn{1}{c|}{}  & \multicolumn{1}{c|}{0.960}              & \multicolumn{1}{c|}{0.060}         & \multicolumn{1}{c|}{0.000}                 & \multicolumn{1}{c|}{} & \multicolumn{1}{c|}{0.968}              & \multicolumn{1}{c|}{0.060}         & \multicolumn{1}{c|}{0.000}                 \\ \cline{2-2} \cline{4-6} \cline{8-10} 
\multicolumn{1}{|c|}{}                & \multicolumn{1}{|c|}{\textbf{Both}}       & \multicolumn{1}{c|}{} & \cellcolor[HTML]{C0C0C0} {1.000}                 & \cellcolor[HTML]{C0C0C0}{0.000}             & \cellcolor[HTML]{C0C0C0}{0.091}             & \multicolumn{1}{|c|}{} & \cellcolor[HTML]{C0C0C0}{1.000}                  & \cellcolor[HTML]{C0C0C0}{0.000}            & \cellcolor[HTML]{C0C0C0}{0.000}                 \\ 
\hline

\end{tabular}
\vspace{-4mm}
\end{table}

\comment{
To compare the result, we trained a random forest classifier with the victim's data using
10 fold cross validation. The trained classifier for these two chosen victims had a FPR
of 4\% and 0\%. When we tested the classifier with an imposter attacker's sample under the 
same settings, the FPR increased to 63.6\% and 50\% respectively.
}

To measure the performance of the imposter in mimicking the victim, we first trained a random forest classifier with the victim's data using
10-fold cross validation.
First, we trained the classifiers with the subset of features that provided the best average results, as 
mentioned in Section \ref{sec:eval}. We analyzed the classifier's accuracy with features
from the phone only, the watch only and both devices. We also trained our classifiers with the
subset of features that provided the best performance for the individual user (victim). Then, we tested these classifiers
against the imposter attacker's data to determine the success rate of the attacker.
The results are shown in Table \ref{tab:ImposterResult}. 

As expected, we found that the individual classifier performed
better than the general classifier.
When the general
classifiers were tested against the imposter attacks, the attacker was able to
imitate the hand motion (captured by watch) of $V_1$ (FPR = 0.909), while he
could not imitate the hip motion (captured by phone) of $V_1$ (FPR = 0.000). On
the other hand, the attacker was able to imitate the hip motion of $V_2$ (FPR =
0.917) while it could not imitate the hand motion of $V_2$ (FPR = 0.000).  When
both devices were used, we can see that the FPR for $V_1$ is low (0.091) but
still high for $V_2$ (0.833). This suggests that the classifier trained with the
features from both devices was dominated by the features from the phone,
and hence the results of impersonation are more similar to that of the phone only.  
Similarly, when the individual best subset features were used to train
the classifier, the attacker could not imitate the
hand motion resulting low attack success rate when 
both devices' features were used.  In other words, \wuzia could
resist the imposters to a high degree when both devices' features and the best
subset of features were used for each individual user.

In summary, these results show that the \wuzia system that leverages both phone
and watch, and employs individualized classifiers can be highly resistant to
walking imitation attacks. This is a significant security advantage of a
multi-device \wuzia scheme.

\vspace{-2mm}
\subsection{Treadmill Attack} \label{sec:treadmill}
\vspace{-1mm}

To perform a more powerful attack on the victim's walking pattern so as to successfully fool the
\wuzia system, we followed the work by Kumar et al. 
\cite{kumar2015treadmill}.
This research represents the state-of-the-art attack against gait biometrics and is 
therefore an ideal platform to evaluate our system against.
In this attack, the attacker already has the sample of a victim's gait pattern.
First, the authors extract different features from the accelerometer sensor of the smartphone to
authenticate users based on their walking pattern to create a baseline model 
called Gait Based Authentication System ($GBAS$). 
Then, they attack on the $GBAS$ system using a treadmill. 
In this attack, instead of imitating the victim's walking pattern, the attacker uses treadmill to control 
different gait characteristics (GCAT) such as speed, step length, step width and thigh lift
to match the features extracted from the victim's walking pattern.
To setup this attack, the attacker first analyzes the feature subsets that dominates
the decision making process of the machine-learning classifiers \cite{kumar2015treadmill}. 
Among these dominant features subset, the attacker then analyzes how these features are
correlated with each other. From this analysis, the attacker tries to 
manipulate only one feature among the correlated features set. Now the attacker
has final set of five features which it needs to manipulate to fool the classifier.
The experimenter creates an imitator profile based on these final five
features mapped to the four GCAT. This mapping is also created using correlation between
GCAT and the dominating feature set. For example, if speed is directly correlated with
the mean of X-axis of the accelerometer ($ACC_{X\_M}$) then to increase or decrease the $ACC_{X\_M}$,
the imitator needs to increase or decrease the walking speed, respectively.

\comment{
Since the \wuzia system extracts different features from different sensors available on the 
smartphone as well as the smartwatch, the attacker will try to analyze which feature subsets
are dominantly selected by the machine-learning classifiers \cite{kumar2015treadmill}. 
Once the attacker learns this, it will try to find which features are correlated with each other.
Then they choose five dominating features which have major impact on the classifier and map
four different gait characteristics (speed, step length, step width, and thigh lift) to these five features.
They use treadmill to control these gait characteristics of the attacker.
}
To thwart such attacks using sophisticated devices like treadmill to control different
gait characteristics, we calculated the correlation values among each pair of features.
The detail regarding the calculation of the correlation among features is explained in Appendix 
\ref{sec:corr} and the results are shown in Figure \ref{fig:correlationAllFeature}. 
From this analysis, we observe that the features 
from the phone are more correlated with the features from the phone while the features from the 
watch are more correlated with the features from the watch.
This means that the attacker cannot use one device to alter the feature of the other device, 
however, it may be able to alter the features from a single device if it knows 
the correlation among the features from the same device.

We next analyzed how the features from a single device are 
correlated with the other features from the same device. The correlations among the features 
from the same device are depicted in Figures \ref{fig:correlationPhoneFeature} and 
\ref{fig:correlationWatchFeature} (Appendix). From these plots, we can see that the features extracted 
from a single sensor were more correlated to each other than the features extracted
from different sensors. For example, mean, standard deviation and range of the accelerometer 
sensor were more correlated with each other, compared to those taken from gyroscope or magnetometer.
We wrote a script to find out the best feature subset such that each feature is correlated
to each other in a given feature subset by less than -/+ 0.1 (i.e., the subset of uncorrelated features).
More the number of uncorrelated features in this subset, harder it will be for the attacker 
to correlate/match all the features with different gait characteristics \cite{kumar2015treadmill}. 
Further one gait characteristics may influence more than one feature vector which do not 
have any correlation, increasing the difficulty of the treadmill attack. 

Further, to increase the performance of the classifier in defending the treadmill
attack, we wrote another script to find out the super set of the subset
containing maximum number of uncorrelated features set. 
The best feature subset for the general classifier in Section \ref{sec:eval} that is trained
with features from both devices consists of eight uncorrelated features. 
This increased the accuracy of the classifier during the benign case while still being robust
to the treadmill attackers.  Further, the treadmill attackers may use more
sophisticated devices to provide better gait characteristics that may alter
different features. We can defend this by increasing the correlation threshold
(currently set to 0.1) for finding uncorrelated feature set.  This will provide
larger number of features that are correlated to each other by that threshold
value. Note that the correlation of 0 to 0.1 is considered near-zero
correlation while that between 0.1 and 0.3 is considered weak correlation
\cite{correlation1,correlation2}.  Hence, using the correlation threshold of
0.3 will still give the feature subset with weak correlation that attacker may
not be able to attack using the treadmill technique.

\vspace{-2mm}
\section{Discussion} \label{sec:discuss}
\vspace{-2mm}
\noindent \textbf{Adherence to Design Criteria}:
Our \wuzia system is compliant with the design goals established in Section \ref{subSec:goals}.
First, \wuzia is triggered and sensor data is polled only when \verifier sends a challenge to
\prover in a challenge-response authentication protocol.  
After \prover has authenticated the user, the system deactivates the sensors.
The classifier model is to be built offline during the training phase.  The
sensor data is collected for no more than 10 seconds and the decision making
process by the classifier is pretty simple (random forest classification).
Hence, \wuzia will have minimal influence on the power consumption and time delay satisfying
our design goals of being lightweight and efficient.

From our results in Table \ref{table:ResultsGeneral}, \wuzia yields very high
F-measure with very low FNR and FPR during the benign case.  The results from
Table \ref{tab:ImposterResult} shows that \wuzia is resistant to imposter
attacks. Further, the use of uncorrelated sensor features makes \wuzia tolerant
to treadmill attacks. This makes \wuzia very robust to errors and attacks.

Last, but not least, \wuzia works in the background while the user walks towards 
\verifier. 
Hence, \wuzia preserves the transparency of \zia even though it adds another layer 
of strong security to the system. 

\smallskip
\noindent
{\bf Fallback Scenarios:} \label{subSec:fallback} 
We showed that our system is very effective with very low FNR. However, a user may be injured, stressed, sick, or 
carrying the phone in a purse or backpack, which may significantly alter the user's walking behavior. 
Such situations can lead to false negatives, as the legitimate user
will be denied access to the system. In such occasional cases, we can fallback to 
traditional password/key based approach for authentication. 

\smallskip
\noindent
{\bf Effect of Changing Apparel or Footwear:} \label{subSec:apparel} 
A user's walking pattern may get affected with the use of varying apparel or footwear.
Our data collection experiment was conducted in lab for a period ranging from 30 to 60 days.
Even though our participants must have worn changing apparel and shoes during the data collection process, 
our classification accuracies are still quite high. This suggests that our classification model
may be robust to changes in walking patterns arising from changing clothing and footwear.

\smallskip
\noindent
{\bf Robotic Attacks:} 
It may be possible to build robots that mimic a user's gait pattern \cite{iop2012robots,klein2012physical}.
For such attacks, the attacker needs to have an access to the 
sample of victim's gait pattern as in case of treadmill attacks \cite{kumar2015treadmill} 
and then program the robots such that the feature
values generated by the sensor with robot's motion matches significantly with that of the 
victim. However, due to the involvement of a robot, it does not seem feasible that such
an attack would be unnoticeable in practice.

\smallskip
\noindent
{\bf Implementing WUZIA on Car Keys:} 
The core idea of \wuzia is not just limited to smartphones. Smart keys were introduced as early 
as 1998 by Mercedes-Benz under the name ``Key-less Go'' 
\cite{keylessGo}. The car keys have evolved from physical keys to Remote Keyless Entry (RKE)
which then led to Passive Keyless Entry (PKE) systems \cite{van2009car}. These keys operate
via RF signals and modern key systems claim that they use encryption to prevent car
thieves from decoding the RF signal \cite{van2009car}.
In 2008, BMW and NXP Semiconductors announced the first multi-functional car key which is
compatible with EMV (Europay, Mastercard, VISA) electronic payment standard. Such keys
contained a dedicated cryptographic coprocessor.
Busold et al. \cite{busold2013smart} introduced smartphone-based NFC-enabled car immobilizers. 
\wuzia can be implemented on any such systems where the key (either physical, RKE or PKE) 
has embedded sensors, processor and RF capability. 

\smallskip \noindent
{\bf Deauthentication:} Our system uses the gait biometrics to authenticate the
user when the user is within the range of the system. 
Our system may also be used to deauthenticate the user.  The traditional
approach to deauthentication relies upon timeouts.
This approach has two major disadvantages. First, a user may get locked when he
is still using the device but inactive for certain duration. Second, the system
may be unlocked for a long duration even when the user has walked away such
that an unauthorized user may get access to the system.
Instead of timeouts, existing \zia systems make use of the RF signal strength
measures to determine whether the prover device has moved away in order to
deauthenticate the user. This approach suffers from a problem that if someone
else (say the user's spouse) takes the prover device away, the user will be
locked out.  \wuzia can effectively address this problem by deauthenticating
the user only when: (1) the RF signals show that the prover device has moved
away, and (2) the system authenticates the user through gait biometrics detection.

\vspace{-2mm}
\section{Related Work on Gait Biometrics} \label{sec:related}
\vspace{-2mm}

The subject of gait biometrics has been well-studied in research literature.
Compared to the existing work, our novelty lies in the use of gait biometrics
for the \zia application, and in the way we extract the gait patterns, i.e.,
using multiple commodity devices and multiple sensors therein. In the rest of this
section, we review the existing literature on gait biometrics.

Many researches have explored the use of accelerometer to
authenticate the users based on their walking pattern. These work mostly use  
electronic motion recording (MR) devices such as MR100 wearable sensor  
\cite{mjaaland2010walk}, ZSTAR \cite{mjaaland2010walk,stang2007gait}, 
ADXL202JQ accelerometers \cite{mantyjarvi2005identifying}, 
MMA7260 \cite{rong2007wearable}, etc. 
These work analyze the accelerometer reading by attaching such MR sensors at 
different location of the body such as waist 
\cite{ailisto2005identifying,mantyjarvi2005identifying,mjaaland2010walk,rong2007wearable} (device wore in a belt), 
lower leg \cite{gafurov2006biometric,sondrol2005using}, 
shoe \cite{chen2008intelligent,huang2007gait,morris2004shoe,yamamoto2008foot}, 
pockets (chest/hip) \cite{gafurov2009gait,vildjiounaite2006unobtrusive}, 
upper limb/forearm \cite{gafurov2009gait}, 
gloves \cite{kim2005new,perng1999acceleration,sama20063did},
and so on. In most of these work, the MR device was tied on the 
particular parts of the body as most of these devices were not wearable.

Vildjiounaite et al. \cite{vildjiounaite2006unobtrusive} used accelerometer module 
(MR sensor) and placed it in chest pocket, hip pocket and hand to authenticate 
users based on their walking pattern. To perform their experiment, they made mock-ups 
of ``clothes with pockets'' from pieces of textile which the users put on over 
their normal clothes. They reported that since it was not the real pocket, shifting 
of the mock-ups of clothes affected the accelerometer readings while the 
accelerometer module itself was not shifting as it was attached to the mock-ups of 
clothes.

Gafurov et al. \cite{gafurov2009gait,gafurov2007spoof} used a ``Motion Recording 
Sensors'' (MRS) to collect accelerometer data.
In their
work \cite{gafurov2007spoof}, they tried to spoof the user's walking pattern by 
performing the experiment in two rounds. First, the targeted user walked in front 
of the attacker twice. Then, the attacker walked alone twice mimicking the user.
They showed that such minimal effort impersonation attack on gait pattern does not increase
the chances of imposters being accepted significantly. In our work, the attacker
watched the victim's walking pattern in person, recorded the pattern in video,
got feedback from his colleagues during training and further got visual feedback
played from the recording of the victim's walking event while authenticating. 
Further they used MRS attached to the belt while we used commercial devices
such as smartphone and smartwatch.

Stang et al. \cite{stang2007gait} also explored the gait based authentication 
approach using ZSTAR accelerometer sensor 
and analyzed if the imposters could imitate the walking pattern. 
They recruited 13 participants to imitate users. Each participant was given 15 attempts on 
each template to attack. The imposters did not see the original walking but they were given a 
simple description of the gait. The participants were provided with the visual 
feedback such that they could see the template gait graph and their gait graph 
continuously plotted on a big screen. The walk duration was 5 second long for each 
walk sample. After each attempt a match score between 0 and 100 was displayed based 
on correlation such that 100 is a perfect match.
They reported 3 persons exceeded the correlation threshold once, 2 persons exceeded the
threshold twice, 1 person exceeded it three times and 1 person managed to exceed as
much as 9 times in 15 attempts.
Therefore, they concluded that it is easy to walk like another person.

Another attempt to mimic walking pattern was made by Mjaaland et al. 
\cite{mjaaland2010walk}. They trained seven imposters to imitate a specific 
victim. They used two wearable sensors: the Motion Recording 100 (MR100), and the
Freescale ZSTAR sensor to record the accelerometer sensor values. They attached
these sensors on belt and asked the participants to wear the belt which could be mounted 
to any person's hip regardless of what they were wearing such that the device
would always have the same-orientation. They conducted short-term hostile scenario
and long-term hostile scenario. In the former scenario, they trained six 
participants for two weeks, five hours every day while in the latter scenario,
they trained the seventh participants for six weeks. In both scenarios, the 
imposters were not able to imitate the victim's walking pattern. They concluded 
that there is a physiologically predetermined boundary to every individual's 
mimicking performance and also that if one successfully adopted gait 
characteristics improved an attacker's performance, other characteristics worsen
in a chain-like effect.

One of the works in line with ours is by Kumar et al. \cite{kumar2015treadmill}
as they also used an Android smartphone with an app to record sensor data
as described in Section \ref{sec:treadmill}.
In this work, they only used features extracted from accelerometer sensors while 
we used features from eight different sensors.
\comment{
They recorded accelerometer sensor only to authenticate users based on their
walking pattern. In this work rather than imitating a victim walking pattern, they 
used different gait characteristics to match the victim's walking pattern.
They assumed that the attacker has the victim's gait samples. This means that the 
attacker knows the values for each feature which would satisfy the classifier 
threshold to authenticate. 
}
From the 47 features extracted from accelerometer sensor only, they
ranked their features based on information gain based attribute evaluator \cite{hall2003benchmarking} and selected 17 top ranked features only.
In our work, we explored the best result for the combination of all 336 feature subset.
\comment{
Since there were many features (47) used by the classifier,
they first selected top 17 ranked features by using information
gain based attribute evaluator \cite{hall2003benchmarking} to build the baseline system.
Then, they analyzed the correlation among 17 selected features and deduced changing one
feature will affect the other correlated features in similar way. With this, they
selected five features which were dominating and could be mapped to four different
gait characteristics. They profiled an imitator by mapping gait characteristics 
to the five dominating features. The imposters were asked to change their gait
characteristics based on this mapping and the victim's feature template so that 
there was a significant match. 
}
Since they were using features extracted from accelerometer sensor only, the features might
be highly correlated and reported that that their system's FAR increased from
5.8\% to 43.66\%. In our work, the best feature subset consists of the eight
uncorrelated features with correlation less than -/+ 0.1. An attacker would
need more sophisticated device than a treadmill to control more gait characteristics
(defined in \ref{sec:treadmill}).

Researchers have also
explored accelerometer and/or gyroscope sensors available on current
smartwatches for the purpose of gait detection. Johnston et al.
\cite{johnston2015smartwatch} used the accelerometer sensor  embedded in the
smartwatch, while Kumar et al.  \cite{kumar2016authenticating} used the
accelerometer and the gyroscope sensor.  In contrast to us, Kumar et al. only
used the sensors from smartwatch and did not consider the use of multiple
devices (both phone and watch).  The authors only extracted a total of 76
features (32 features from the accelerometer readings and 44 features from the
gyroscope readings), while we work with a total of 336 features, resulting in
much lower FNR and FPR.
Also, unlike our work, Kumar et al. did not study active attacks
and only reported the performance under the zero-effort or random attack.
Moreover, the targeted applications for the two works are different (\zia vs.
continuous authentication).

\vspace{-3mm}
\section{Conclusion and Future Work} \label{sec:conclusion}
\vspace{-3mm}
We proposed the use of walking-based biometrics to protect
zero-interaction authentication systems in the event of loss or theft of
authentication tokens.  Our approach transparently authenticates the user to
her authentication token as she walks towards the authentication terminal in
order to unlock it. Our system leverages a smartphone and/or a
smartwatch, and multiple embedded sensors therein, to reliably detect the
unique walking pattern of the user.  Our results suggest that especially when
using both devices together, the system offers almost error-free detection and makes
it very difficult for even a powerful attacker to imitate a user's walking habit.
Consequently, we believe that our approach can significantly enhance
the security of current zero-interaction systems without degrading their
usability. 

Future work may explore other types of wearable devices (such as glasses, which
may capture head movements, or shoes, which may capture feet movements) to
further extend our approach, study the implementation of similar techniques on
car keys in keyless entry systems, and conduct broader data collection
campaigns with larger and diverse population samples.

\vspace{-2mm}
\bibliographystyle{splncs03}
{ \small
\bibliography{main-paper}
}

\appendix
\section*{Appendix}

\vspace{-2mm}

\section{Correlation Analysis} \label{sec:corr}
\vspace{-2mm}
Correlation is commonly used to find the relationship between two or more objects.
To find the similarity between two different features, we calculated the correlation as 
follows. Let x and y be the values of two feature vectors and we have $n$ data samples for 
each feature vector. 

{\scriptsize{
\begin{equation}
S_{xx} = \sum x^{2} - \frac{(\sum x)^2}{n}; \hspace{5mm} 
S_{yy} = \sum y^{2} - \frac{(\sum y)^2}{n} 
\end{equation}

\begin{equation}
S_{xy} = \sum xy - \frac{(\sum x) (\sum y)}{n}
\end{equation}
}}

The correlation ($\sigma$) between the features vectors x and y  is

{\scriptsize{
\begin{equation}
\sigma = \frac{S_{xy}}{\sqrt{S_{xx}S_{yy}}}
\end{equation}
}}

The calculation of the correlation ($\sigma$) is based on the alternative computation formula 
for Pearson's r \cite{pearsonR}. The alternative computation for the Pearson's r avoids
the step of computing deviation scores.

\begin{figure}[t]
\centering
  \includegraphics[width=.8\columnwidth]{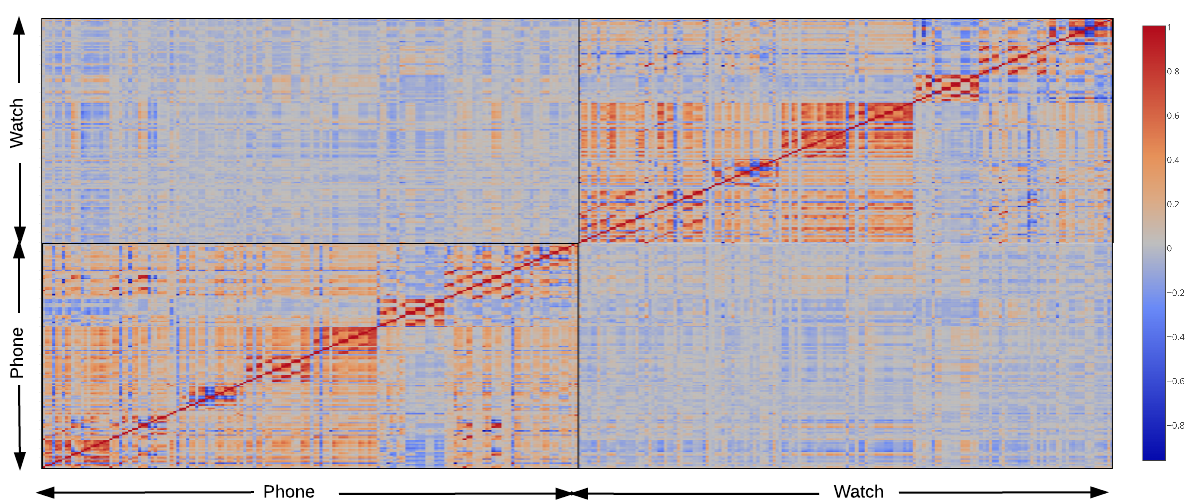}
 \vspace{-3mm}
 \caption{\scriptsize Heatmap showing pairwise correlation values between all the 336 features. 
 Red depicts high positive correlation, Blue depicts high negative correlation, and
 Grey depicts low correlation. The first half (first half rows 
 and the first half columns) represents the features associated with the phone 
 while the second half (the second half rows and the second half columns)  
 represents the features associated with the watch. We can see that the features of phones 
 are more correlated with those of phone while features of watch are more correlated with 
 those of watch.}
 \label{fig:correlationAllFeature}
 \vspace{-5mm}
\end{figure}

\begin{figure}[t]
\centering
  \includegraphics[width=.8\columnwidth]{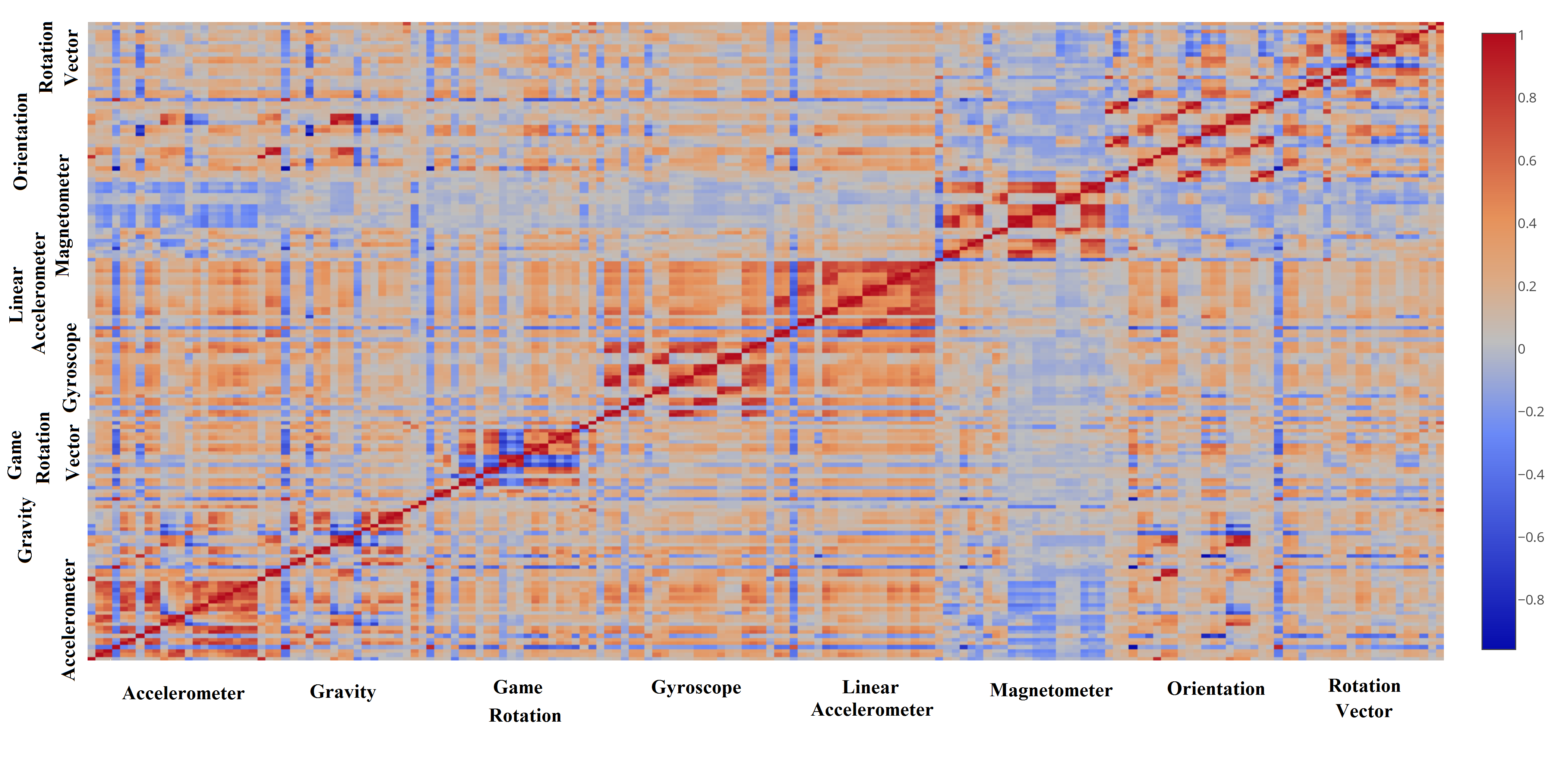}
  \vspace{-6mm}
\caption{\scriptsize Heatmap showing pairwise correlation between the features from phone only. 
The color depiction is same as in previous figure. The features extracted using the same sensor have higher correlation
 compared to those extracted using different sensors. 
 }
 \label{fig:correlationPhoneFeature}
 \vspace{-6mm}
\end{figure}

\begin{figure}[t]
\centering
  \includegraphics[width=.8\columnwidth]{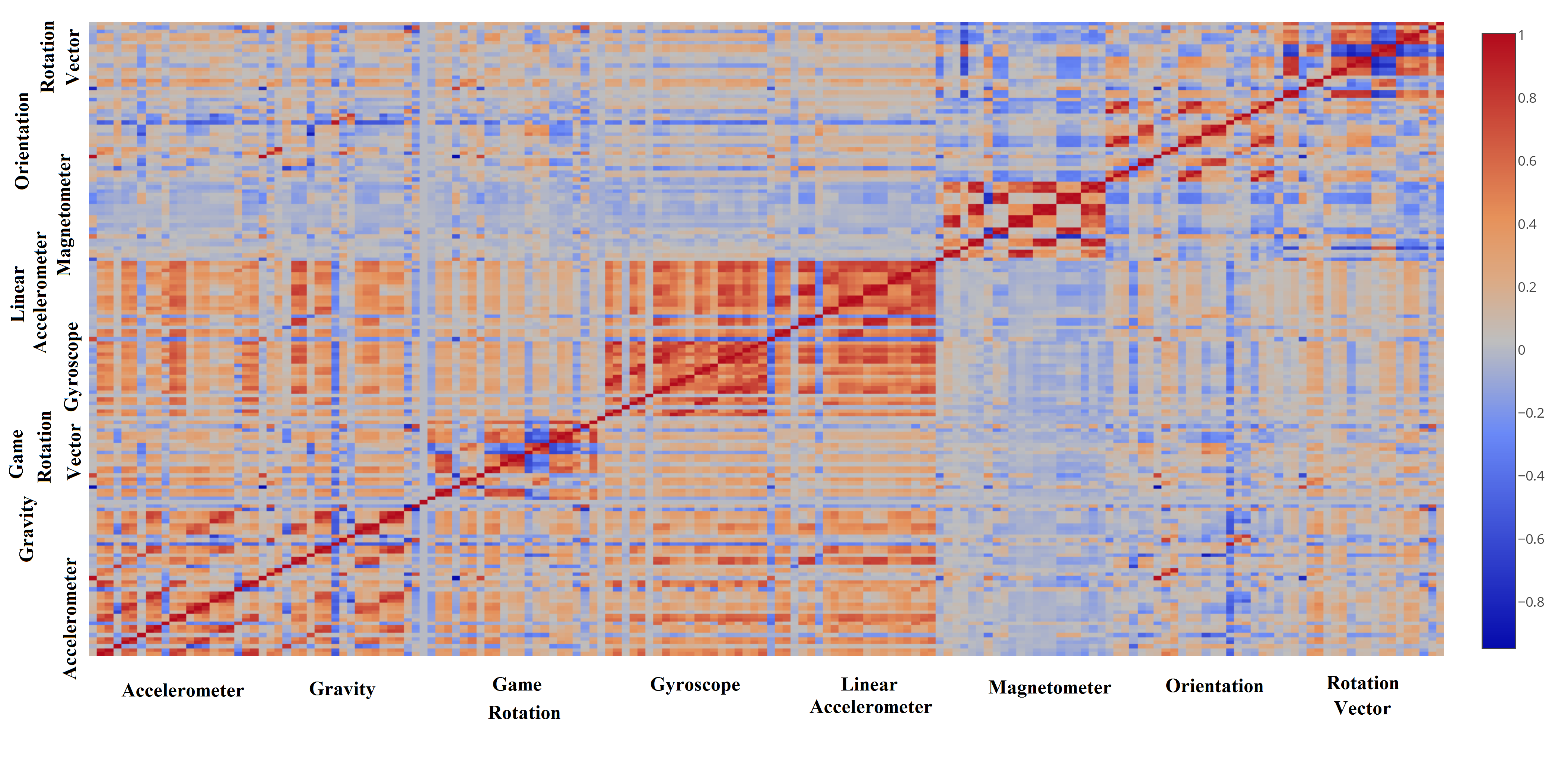}
  \vspace{-6mm}
 \caption{\scriptsize Heatmap showing pairwise correlation between the features from watch only. 
The color depiction is same as in previous figure. The features extracted using the same sensor have higher correlation
 compared to those extracted using different sensors. 
 }
 \label{fig:correlationWatchFeature}
 \vspace{-6mm}
\end{figure}

\end{document}